\documentclass[aps,prd,reprint,amsmath,amssymb]{revtex4-2}

\usepackage{graphicx}
\usepackage{dcolumn}
\usepackage{bm}
\usepackage{siunitx}
\usepackage[mathlines]{lineno}
\usepackage{graphicx}
\usepackage{txfonts}
\usepackage{hyperref}
\hypersetup{
    colorlinks=true,
    linkcolor=blue,
    urlcolor=blue,
    citecolor=blue,
    }

\begin{document}

\title{Revealing the internal magnetic fields of magnetars via their associated periodic signals}


\author{Jie Shu}
    \affiliation{Institute of Astrophysics, Central China Normal University, Wuhan 430079, China}

\author{Quan Cheng}
    \email{qcheng@ccnu.edu.cn}
    \affiliation{Institute of Astrophysics, Central China Normal University, Wuhan 430079, China}
\author{Xiao-Ping Zheng}
    \affiliation{Institute of Astrophysics, Central China Normal University, Wuhan 430079, China}

\date{\today}

\begin{abstract}
The magnetic deformation of magnetars is affected by their internal magnetic fields, which are generally difficult to be measured directly through observations. In this work, the  periodic pulse-phase modulations in the hard X-ray emissions of the magnetars 4U 0142+61, 1E 1547.0-5408, SGR 1900+14, and SGR 1806-20, and the periodicities of fast radio bursts (FRBs) 180916 and 121102 are interpreted as free precession of the (host) magnetars. Using these periodic signals, we investigate the magnetars’ internal magnetic fields. In order to simultaneously account for the modulation periods and surface thermal emissions of the former four magnetars, and require that their internal poloidal fields smoothly connect with the surface dipole fields, the parameter that characterizes the distribution of toroidal field in the magnetar interior should satisfy $\beta\gtrsim1$. Moreover, their volume-averaged strengths of poloidal and toroidal fields are respectively $\bar{B}_{\rm p}\sim10^{14}$--$10^{15}$ G and $\bar{B}_{\rm t}\sim10^{15}$ G with the strength ratios $\bar{B}_{\rm t}/\bar{B}_{\rm p}$ generally distributing within $\sim2$--$37$. We could also constrain the critical temperature for neutron superfluidity in the neutron-star core considering that the former four magnetars are probably precessing, and the most stringent constraint is $T_{\rm c,core}<6.4\times10^8$ K. Adopting a possible critical temperature $T_{\rm c,core}=5\times10^8$ K, we could obtain $\bar{B}_{\rm p}\gtrsim10^{14}$--$10^{15}$ G and $\bar{B}_{\rm t}\gtrsim10^{14}$--$10^{15}$ G for the host magnetars of FRBs 180916 and 121102, which indicates that the magnetars of our interest possibly have similar poloidal and toroidal fields.
\end{abstract}

\maketitle

\section{Introduction}\label{Sec I}

Magnetars are generally regarded as a special kind of neutron stars (NSs) that possess ultra-strong dipole magnetic fields with a typical strength of $\sim10^{14}$--$10^{15}$ G, however, slow spins with a typical period of $\sim1$--$12$ s \cite{2014ApJS..212....6O}. They are usually observed as soft gamma-ray repeaters (SGRs) and/or anomalous X-ray pulsars (AXPs) \cite{2014ApJS..212....6O}, and these energetic emissions are probably powered by the strong magnetic fields \cite{1982ApJ...260..371K,1992ApJ...392L...9D,1996ApJ...473..322T}. Moreover, recent observations showed that magnetars can be the host of at least some of the fast radio bursts (FRBs) \cite{2020Natur.587...54C,2020Natur.587...59B}, a class of brilliant millisecond-duration radio signals of extragalactic origin \cite{2007Sci...318..777L,2013Sci...341...53T,2019A&ARv..27....4P}. The strong magnetic fields may be responsible for the FRBs and associated X-ray bursts as observed in SGR J1935+2154/FRB 200428 \cite{2020Natur.587...59B,2014MNRAS.442L...9L,2021NatAs...5..378L}. The origin of magnetars' strong magnetic fields is a long-standing puzzle since their discovery (see, e.g., \cite{2017ARA&A..55..261K}). It is widely accepted that their strong magnetic fields may originate from the fossil fields of progenitors or the dynamo processes during the formation of NSs \cite{1992ApJ...392L...9D,2006MNRAS.367.1323F}. 

Though the dipole fields of magnetars can be estimated from the measured spin periods $P$ and their first derivatives $\dot{P}$, little information about the strengths and configurations of internal magnetic fields of magnetars is known currently. It is possible that their internal magnetic fields have a twisted-torus shape consist of both poloidal and toroidal components \cite{2004Natur.431..819B,2006A&A...450.1097B}. However, which component of the internal fields plays a dominant role in distorting the magnetars is still an open issue (e.g., \cite{2010MNRAS.406.2540C,2011MNRAS.417.2288M}). The internal field configuration and energies of the poloidal and toroidal fields generally determines the shapes of the deformed magnetars \cite{2010MNRAS.406.2540C,2011MNRAS.417.2288M,2015ApJ...798...25D}. Specifically, the magnetars will be a prolate (oblate) ellipsoid and have magnetically-induced ellipticity $\epsilon_{\rm B}<0$ ($\epsilon_{\rm B}>0$) if the toroidal (poloidal) fields are dominant \cite{2002PhRvD..66h4025C,2009MNRAS.398.1869D}. Once the magnetic and spin axes of the magnetars with $\epsilon_{\rm B}<0$ are not orthogonal, namely the magnetic tilt angle between the two axes satisfies $\chi\neq\pi/2$, the magnetic axis will freely precess around the spin axis \cite{2002PhRvD..66h4025C,2005ApJ...634L.165S,2019PhRvD..99h3011C,2024SCPMA..67.129513}, resulting in a periodic modulation of the spins of the magnetars. In contrast, for the magnetars with $\epsilon_{\rm B}>0$, if their tilt angles are $\chi\neq0$ (the magnetic axis is not aligned with the spin axis), free precession of the magnetars will also occur \cite{2009MNRAS.398.1869D,2019PhRvD..99h3011C,2024SCPMA..67.129513}.  

Free precession of magnetars can be used to account for the observed periodic pulse-phase modulations in the hard X-ray emissions of the magnetars 4U 0142+61, 1E 1547.0-5408, SGR 1900+14, and SGR 1806-20 \cite{2014PhRvL.112q1102M,2021MNRAS.502.2266M,2021ApJ...923...63M,2024PASJ...76..688M}, the periodic modulations in the light curves of some gamma-ray bursts \cite{2021MNRAS.502.2482S,2024ApJ...973..126Z}, as well as the periodicities of FRBs 180916.J0158+65 (hereafter 180916) \cite{2020Natur.582..351C,2020ApJ...895L..30L,2020ApJ...892L..15Z,2022ApJ...928...53W} and 121102 \cite{2020MNRAS.495.3551R,2021MNRAS.500..448C,2022ApJ...928...53W,2024ApJ...969...23L}, especially when considering that FRB 200428 indeed originated from the magnetar SGR J1935+2154 \cite{2020Natur.587...54C,2020Natur.587...59B,2021NatAs...5..378L}. Although other models including the forced precession of rotating magnetars caused by external electromagnetic torques \cite{2020MNRAS.497.1001S}, forced precession of the NS by a fallback disk \cite{2020RAA....20..142T}, orbit-induced spin precession of the NS in a compact binary \cite{2020ApJ...893L..31Y}, NS accretor in a NS-white dwarf binary \cite{2020MNRAS.497.1543G}, and Lense-Thirring precession of the debris disk around a NS \cite{2020PASJ...72L...8C}, may also account for the periodicities of FRBs 180916 and 121102, we mainly focus on the scenario of freely precessing magnetars. Desvignes et al. \cite{2024NatAs...8..617D} proposed that the scenario of freely precessing magnetars may not explain the periodicities of repeating FRBs as the precession of the magnetar XTE J1810-197 was observed to be damped on a timescale of months. However, more observational evidences are needed before reaching a definite conclusion given that similar damping features were not observed in 4U 0142+61, 1E 1547.0-5408, SGR 1900+14, and SGR 1806-20, which also showed evidences of free precession \cite{2014PhRvL.112q1102M,2021MNRAS.502.2266M,2021ApJ...923...63M,2024PASJ...76..688M}.   

In this work, six sources that showed periodic emissions are investigated, which are 4U 0142+61, 1E 1547.0-5408, SGR 1900+14, SGR 1806-20, and FRBs 180916 and 121102. The first four sources are explicitly magnetars which exhibited periodic modulations in their hard X-ray emissions as introduced above. The modulation periods $P_{\rm m}$ are summarized in Tab. \ref{tab1} in unit of kiloseconds (ks). For simplicity, the error bars in $P_{\rm m}$ are neglected in our calculations. The same as in Refs. \cite{2014PhRvL.112q1102M,2021MNRAS.502.2266M,2021ApJ...923...63M,2024PASJ...76..688M}, $P_{\rm m}$ are regarded as the free-precession periods $P_{\rm p}$ of the magnetars, i.e., $P_{\rm m}=P_{\rm p}$. The last two are repeating FRBs that possibly associated with magnetars. Since how the FRBs are formed from magnetars and the emission region of FRBs are still open questions (e.g., \cite{2019PhR...821....1P,2020MNRAS.496.3390B,2024arXiv240705366B}), these issues are beyond the scope of this work. We only intend to illustrate the periodicities of FRBs 180916 and 121102 as the precession periods of their putative host magnetars as that done in \cite{2020ApJ...895L..30L,2020ApJ...892L..15Z,2022ApJ...928...53W}. For FRBs 180916 and 121102, we therefore respectively have $P_{\rm p}\simeq16.35$ and 160 d \cite{2020Natur.582..351C,2021MNRAS.500..448C}.  

To interpret these periodic signals in the scenario of freely precessing magnetars, the precession of magnetars should last for long enough ($\gg P_{\rm p}$), and thus not be damped by internal viscous dissipation. As discussed in \cite{2020ApJ...895L..30L}, one critical condition for this to be realized is that the neutrons in the NS core are \textit{not} in the superfluid phase. In other words, the NS internal temperature should be above the critical temperature for the occurrence of neutron superfluidity in the core since free precession of magnetars could be quickly damped due to mutual frictions between superfluid neutrons and relativistic electrons \cite{1988ApJ...327..723A,2002PhRvD..66h4025C,2009MNRAS.398.1869D}, and even stronger mutual interactions between neutron vortices and flux tubes in the NS core \cite{2003PhRvL..91j1101L}. In fact, high enough internal temperatures of $\sim10^8$ K may be sustained for $\sim10^5$ yr because of ambipolar diffusion \cite{1992ApJ...395..250G} of the strong magnetic fields in the magnetar core (e.g., \cite{1996ApJ...473..322T,2009MNRAS.398.1869D,2012MNRAS.422.2878D}). Such internal temperatures could be above the critical temperature for $^3P_2$ neutron superfluidity in the core (see the discussions in Sec. \ref{Sec III}). It is natural that both the poloidal and the toroidal fields can participate in the dissipation process and heat the magnetars. Meanwhile, in the freely-precessing-magnetar scenario, by using $P_{\rm p}$ and spin periods of the magnetars, one can estimate their ellipticities, which are contributed by both poloidal and toroidal fields in the core (e.g., \cite{2002PhRvD..66h4025C,2011MNRAS.417.2288M,2015ApJ...798...25D}). We thus envisage that by combining the requirements on the internal temperatures and ellipticities of the magnetars, constraints on their poloidal and toroidal fields may be set.

This work is organized as follows. Free precession of magnetars is introduced in Sec. \ref{Sec II}. The relation between the internal fields and internal temperatures of magnetars is given in Sec. \ref{Sec III}. Constraints on the poloidal and toroidal fields of the magnetars are presented in Sec. \ref{Sec IV}. Finally, conclusions and discussions are given in Sec. \ref{Sec V}.
 
\section{Free precession of magnetars}\label{Sec II}

Generally, if the magnetic and spin axes of a rotating magnetically deformed magnetar are neither aligned ($\chi\neq0$) nor orthogonal ($\chi\neq\pi/2$), free precession of the magnetar will naturally occur since its spin energy can be minimized during precession \cite{2009MNRAS.398.1869D,2019PhRvD..99h3011C,2024SCPMA..67.129513}. The precession period approximately takes the form
\begin{equation}
\begin{aligned}
    P_{\rm p}\simeq P/|\epsilon_{\rm B}|,
\end{aligned}
\label{Pp}
\end{equation}
where $P$ is the magnetar's spin period. As introduced in Sec. \ref{Sec I}, $\epsilon_{\rm B}$ can be either positive or negative depending on whether the poloidal or the toroidal component is dominant, and $\epsilon_{\rm B}$ in general is contributed by the two components. Assuming that the NS is incompressible, non-superconducting, and has a uniform density and both poloidal and toroidal fields in its interior, the magnetically-induced ellipticity $\epsilon_{\rm B}$ was obtained in \cite{1969ApJ...157.1395O} (see also \cite{2002PhRvD..66h4025C}). Here, the basic form of the expression for $\epsilon_{\rm B}$ is used as it explicitly presents the contributions from poloidal and toroidal fields \cite{2002PhRvD..66h4025C}. However, we slightly revised the expression to the following form  
\begin{equation}
\begin{aligned}
    \epsilon_{\rm B}=\frac{25R^4}{24GM^2}\left(\alpha \bar{B}_{\rm p}^2-\beta \bar{B}_{\rm t}^2 \right),
\end{aligned}
\label{epsilonB}
\end{equation}
where $R$ and $M$ are respectively the NS's radius and mass, and $R=12$ km and $M=1.4M_\odot$ are adopted in the calculations. $\bar{B}_{\rm p}$ and $\bar{B}_{\rm t}$ are the volume-averaged strengths of the poloidal and toroidal fields, respectively. Following Ref. \cite{2002PhRvD..66h4025C}, we take $\alpha=21/10$. The coefficient $\beta$ accounts for the effect of toroidal-field distribution on the NS's deformation, and is required to satisfy $\beta\geq1$. Obviously, Eq. (\ref{epsilonB}) returns to the original form (Eq. (2.5)) given in \cite{2002PhRvD..66h4025C} when $\beta=1$ is taken, which just corresponds to the case of uniformly-distributed toroidal field \cite{1969ApJ...157.1395O}. Since the toroidal field may actually be confined in a certain internal region along the equator \cite{2004Natur.431..819B,2006A&A...450.1097B,2011MNRAS.417.2288M,2015ApJ...798...25D}, rather than the whole volume of the NS \cite{1969ApJ...157.1395O}, in this case we probably have $\beta>1$. Our prescription for $\epsilon_{\rm B}$ is physically reasonable because keeping the energy of the toroidal field unchanged, the ellipticity induced by the toroidal field would be larger if it is confined in a certain region around the equatorial plane inside the NS, in comparison with the case of uniformly-distributed toroidal field. We note that the effect of toroidal-field distribution on $\epsilon_{\rm B}$ has been found in the case when core superconductivity is involved \cite{2012MNRAS.421..760M}.

Free precession of the magnetar can be damped due to internal viscous dissipation on a timescale $\tau_{\rm dis}\simeq\xi P/|\epsilon_{\rm B}|$ \cite{2002PhRvD..66h4025C}, where $\xi$ is the number of precession cycles. The value of $\xi$ depends on specific viscous mechanisms that lead to dissipation of the stellar precessional energy \cite{2002PhRvD..66h4025C,2019PhRvD..99h3011C,2024SCPMA..67.129513}. Generally, if the magnetar's precessional energy is dissipated through scattering between superfluid neutrons and relativistic electrons in the NS core, $\xi$ is within $10^2$--$10^4$ \cite{1988ApJ...327..723A}. Therefore, the damping timescale of the magnetar's free precession can be estimated as $\tau_{\rm dis}\simeq10^{10}\xi_{4} P_{0}/|\epsilon_{\rm B}|_{-6}~{\rm s}$. In this work, we use the notation $A_{x}=A/10^{x}$. Obviously, free precession of the magnetar could be damped in $\lesssim300$ yr, which is less than or comparable to the characteristic ages $\tau_{\rm c}$ \footnote{The damping timescale is much shorter than the kinematic ages $\tau_{\rm k}$ measured from proper motions of SGR 1900+14 and SGR 1806-20 as listed in Tab. \ref{tab1}.} of the magnetars as presented in Tab. \ref{tab1}. As a result, to account for the periodic emissions associated with these magnetars in the precessing-magnetar scenario, it is required that the viscous mechanism above cannot work. To avoid damping of the magnetars' free precession, the neutrons in their cores cannot be superfluid, thus their core temperatures should be above the critical value for $^3P_2$ neutron superfluidity in the core. 

\section{Relation between the internal fields and internal temperatures of magnetars}\label{Sec III}

The internal region below the heat-blanketing envelope (with a density $\rho\gtrsim10^{9}~{\rm g}~{\rm cm}^{-3}$) of the magnetar is approximately isothermal due to the high thermal conductivity \cite{2003ApJ...594..404P,2007Ap&SS.308..353P,2016ApJ...833..261B}. Thus the temperature in the region with $\rho\gtrsim10^{9}~{\rm g}~{\rm cm}^{-3}$ of the magnetar may represent the temperature of its core. Denoting the temperature in this region as $T_{\rm b}$, it can be inferred from the surface temperature $T_{\rm s}$ of the magnetar after considering the effects of the chemical composition, anisotropic thermal conductivity caused by strong magnetic fields, and neutrino emission in the envelope, as well as the state of matter at the stellar surface \cite{2003ApJ...594..404P,2007Ap&SS.308..353P,2016ApJ...833..261B}. Specifically, whether the magnetar has a gaseous atmosphere or a condensed surface cannot obviously affect the $T_{\rm b}$--$T_{\rm s}$ relation \cite{2007Ap&SS.308..353P,2016ApJ...833..261B}, and the neutrino emission in the envelope becomes effective only when $T_{\rm b}\gtrsim10^9$ K, which is higher than the inferred $T_{\rm b}$ (see Tab. \ref{tab1}) of the magnetars of interest here. Therefore, these two effects can be neglected, the chemical composition and magnetically-induced anisotropy in thermal conductivity of the envelope could be important in determining $T_{\rm b}$ from the observed $T_{\rm s}$ of the magnetars. 

Following Ref. \cite{2003ApJ...594..404P}, for a non-magnetized envelope that consists of iron, $T_{\rm b}$ can be inferred from $T_{\rm s}$ of the NS by using 
\begin{equation}
\begin{split}     
&T_{\text{s6}}^{4}(0)=g_{\text{14}}\left[(7\zeta)^{2.25}+(0.33\zeta)^{1.25}\right],\\
    {\rm with}~~ &\zeta=T_{\text{b9}}-10^{-3}g_{\text{14}}^{1/4}\sqrt{7T_{\text{b9}}}.
\end{split}
\label{Ts1}
\end{equation}
In the above formula, $g_{\rm 14}$ is the gravitational acceleration at the NS surface in unit of $10^{14}~{\rm cm}~{\rm s}^{-2}$. The ``$0$" in the parenthesis in Eq. (\ref{Ts1}) denotes $B=0$ (non-magnetized). The strong magnetic field $B$ in the envelope can enhance heat transport along the field line, however, impede it in the transverse direction of the field line. In the case of a magnetized iron envelope, the NS's surface temperature is expressed as \cite{2003ApJ...594..404P}
\begin{equation}
    T_{\text{s6}}(B)=T_{\text{s6}}(0)X,
\label{Ts2}
\end{equation}
where $X$ is the magnetic correction factor depending on $B$, $T_{\rm b}$, and the angle $\theta$ between the field line and the normal to the stellar surface. The specific form of $X$ reads \cite{2003ApJ...594..404P}
\begin{equation}
\begin{split}
&X=\left(X_{\parallel}^{\alpha}\cos^{2}\theta+X_{\bot}^{\alpha}\sin^{2}\theta\right)^{1/\alpha},\\
{\rm with}~~&\alpha=4+\sqrt{X_{\bot}/X_{\parallel}}.
\end{split}
\label{X}
\end{equation}
The expressions for $X_{\parallel}$ and $X_{\bot}$ are given by Eqs. (A5) and (A6) in \cite{2003ApJ...594..404P}. 

Assuming that 4U 0142+61, 1E 1547.0-5408, SGR 1900+14, and SGR 1806-20 have an iron envelope, and in the envelope the magnetic fields predominantly have a dipole form with their strengths being approximately equal to the surface dipole fields, then from Eqs. (\ref{Ts1})-(\ref{X}) we can estimate $T_{\rm b}$ by using the measured $T_{\rm s}$ (see Tab. \ref{tab1} and the reference therein) of these magnetars and setting $\theta\approx0$. The inferred $T_{\rm b}$ of the magnetars are presented in Tab. \ref{tab1}, which will later be used to obtain their $\bar{B}_{\rm p}$ and $\bar{B}_{\rm t}$. For FRBs 180916 and 121102, since their putative host magnetars have not been discovered, both $T_{\rm s}$ and $P$ are definitely unavailable. Nevertheless, if $P$ are assumed for the host magnetars, by combining $P$, $P_{\rm p}$, and the critical temperature for $^3P_2$ neutron superfluidity in the NS core, $T_{\rm c,core}$, we may also constrain their $\bar{B}_{\rm p}$ and $\bar{B}_{\rm t}$.

From the aforementioned discussions we realize that to avoid damping of free precession of the magnetars, their internal temperatures $T_{\rm b}$ should satisfy $T_{\rm b}>T_{\rm c,core}$. Generally, $T_{\rm b}$ of a magnetar can be determined by solving the thermal evolution equation, in which the cooling due to neutrino emission from modified Urca processes in the core and thermal emission from the stellar surface, and the heating arising from dissipation of the internal magnetic fields are all involved (e.g., \cite{1998ApJ...506L..61H,2003ApJ...594..404P,2006NuPhA.777..497P,2016ApJ...833..261B}). The modified Urca neutrino emission can generally dominate over the surface thermal emission in the cooling of the magnetars at current ages. When the former is offset by heating from ambipolar diffusion of the core magnetic fields \cite{1992ApJ...395..250G,1996ApJ...473..322T,2016ApJ...833..261B,2012MNRAS.422.2878D}, namely, thermal equilibrium between neutrino cooling and heating from magnetic field decay is established, $T_{\rm b}$ of the magnetar can be expressed as \cite{2016ApJ...833..261B}  
\begin{equation}
\begin{split}
    T_{\rm b}&\simeq 8\times 10^8\left(\frac{\bar{B}_{\rm p,16} \delta \bar{B}_{\rm p,16}}{L_{5}}\right)^{0.2} \left(\frac{\rho_{\rm c}}{\rho_{\rm nuc}}\right)^{-7/30}\\
    &+8\times 10^8\left(\frac{\bar{B}_{\rm t,16} \delta \bar{B}_{\rm t,16}}{L_{5}}\right)^{0.2} \left(\frac{\rho_{\rm c}}{\rho_{\rm nuc}}\right)^{-7/30}~{\rm K}, 
\end{split}
\label{Tb}
\end{equation}
where $\delta \bar{B}_{\rm p}$ and $\delta \bar{B}_{\rm t}$ are the variations in $\bar{B}_{\rm p}$ and $\bar{B}_{\rm t}$, respectively. Since $\delta \bar{B}_{\rm p}$ and $\bar{B}_{\rm p}$ could be comparable and so are $\delta \bar{B}_{\rm t}$ and $\bar{B}_{\rm t}$ \cite{2016ApJ...833..261B}, we adopt $\delta \bar{B}_{\rm p,16}/\bar{B}_{\rm p,16}=\delta \bar{B}_{\rm t,16}/\bar{B}_{\rm t,16}=0.5$. For the length scale $L$ on which the magnetic fields vary, we take $L_{5}=1$ \cite{1992ApJ...395..250G,1998ApJ...506L..61H}. $\rho_{\rm c}$ and $\rho_{\rm nuc}$ are respectively the density of the core and nuclear saturation density, and $\rho_{\rm c}/\rho_{\rm nuc}=1$ is used for simplicity.  

\begin{table*}[htbp]
   \centering
   \caption{\centering{The spin period $P$, modulation period $P_{\rm m}$, magnetically-induced ellipticity $\epsilon_{\rm B}$, first derivative of the spin period $\dot{P}$, surface dipole field at the magnetic pole $B_{\rm s}$, characteristic age $\tau_{\rm c}$, kinematic age $\tau_{\rm k}$, surface temperature $T_{\rm s}$, internal temperature $T_{\rm b}$, and strength ratio of toroidal to poloidal fields $\bar{B}_{\rm t}/\bar{B}_{\rm p}$ when the two requirements (see the text) are satisfied for the magnetars 4U 0142+61, 1E 1547.0-5408, SGR 1900+14, and SGR 1806-20.}}
   \begin{tabular}{ccccccccccc}
   \hline
   \hline
   \textbf{Magnetars} &  $P$~(s)\footnote{Data from \href{http://www.physics.mcgill.ca/~pulsar/magnetar/main.html}{the McGill Online Magnetar Catalog \cite{2014ApJS..212....6O}: http://www.physics.mcgill.ca/~pulsar/magnetar/main.html}} & $P_{\rm {m}}~(\unit{ks})$ &  $\epsilon_{\rm B}~(10^{-4})$   & $\dot{P}~({10^{-11}~ {\rm s/s}})\footnotemark[1]$  &  $B_{\rm s}~(10^{14}~\unit{G})$\footnotemark[1]  &  $\tau_{\rm {c}}~(\unit{kyr}) $\footnotemark[1] & $\tau_{\rm {k}}~(\unit{kyr})$\footnote{Vigan$\grave{\unit{o}}$ et al. \cite{2013MNRAS.434..123V}} & $T_{\rm s}~(10^6~\unit{K})$ \footnotemark[1]  & $T_{\rm b}~(10^8~\unit{K})$   &  $\bar{B}_{\rm t}/\bar{B}_{\rm p}$\\
   
   \hline
    \textbf{4U 0142+61}     & 8.69  & 55 $\pm$ 4  \footnote{Makishima et al.  \cite{2014PhRvL.112q1102M}} & -1.58 &  0.20  & 1.3  & 68.00   &     -     &  3.17  &  6.44  & 37.28 (when $\beta\simeq1.6$)  \\
    \textbf{1E 1547.0-5408} & 2.09  & 36.0 $\pm$ 2.3  \footnote{Makishima et al.  \cite{2021MNRAS.502.2266M}} & -0.58 &  4.77  & 3.2  &  0.69   &     -     &  3.33  &  6.49  & 11.90 (when $\beta\simeq1.0$)  \\
    \textbf{SGR 1900+14}    & 5.20  & 40.5 $\pm$ 0.8  \footnote{Makishima et al.  \cite{2021ApJ...923...63M}} & -1.28 &  9.20  & 7.0  &  0.90   & 3.98-7.94 &  3.64  &  7.15  &  5.48 (when $\beta\simeq2.2$)  \\
    \textbf{SGR 1806-20}    & 7.55  & 16.44 $\pm$ 0.02 \footnote{Makishima et al.  \cite{2024PASJ...76..688M}} & -4.59 & 49.50  & 20.0 &  0.24   & 0.63-1.00 &  4.26  &  8.42  &  1.93 (when $\beta\simeq7.9$)  \\
   \hline
   \end{tabular}
   \label{tab1}
\end{table*}

The critical temperature $T_{\rm c,core}$ for $^3P_2$ neutron superfluidity still remains to be an open question because it can be affected by both complicated interactions between nucleons and medium effects \cite{1998PhRvC..58.1921B,2004PhRvL..92h2501S,2004PhRvL..93o1101K,2023ApJ...955...76K}, and may distribute in a rather wide range below $\sim10^{10}$ K \cite{2004ApJS..155..623P}. Observations of the surface thermal emissions of some isolated NSs suggested that $T_{\rm c,core}$ is possibly within $\sim10^{8}$--$10^{9}$ K \cite{2011PhRvL.106h1101P,2018PhRvC..97a5804B}. Such a range for $T_{\rm c,core}$ is also supported by some theoretical calculations, which take into account the effect of nucleon-nucleon short-range correlation \cite{2016PhRvC..94b5802D,2021MNRAS.500.1505D}. Therefore, $T_{\rm c,core}\sim10^{8}$--$10^{9}$ K is assumed in this work.  

In Tab. \ref{tab1}, we also present the spin periods $P$, their first time derivatives $\dot{P}$, and surface dipole fields at the magnetic pole $B_{\rm s}$ of the four magnetars \cite{2014ApJS..212....6O}, all of which will be used in the analysis below. In view of the FRB 200428 and SGR J1935+2154 association \cite{2020Natur.587...54C,2020Natur.587...59B,2021NatAs...5..378L}, $P$ of the putative host magnetars of FRBs 180916 and 121102 are assumed to be the same as typical Galactic magnetars. Thus three typical values $P=1$, $5$, and $10$ s are adopted for the host magnetars. It should be noted that by using $P$ and $P_{\rm p}$ of the four magnetars, one can only obtain $|\epsilon_{\rm B}|$ from Eq. (\ref{Pp}) without knowledge of the shapes of the deformed magnetars (also the signs of $\epsilon_{\rm B}$). However, as proposed in \cite{2014PhRvL.112q1102M}, these magnetars are more likely to have a prolate shape (which actually corresponds to $\epsilon_{\rm B}<0$ \cite{2002PhRvD..66h4025C,2009MNRAS.398.1869D}) since the wobbling angles of nearly all magnetars are non-zero. The values for $\epsilon_{\rm B}$ of the four magnetars are listed in Tab. \ref{tab1}. As no observations could be used to infer the shapes of the host magnetars of FRBs 180916 and 121102, both $\epsilon_{\rm B}>0$ and $\epsilon_{\rm B}<0$ will be considered in our analysis.   

\begin{figure*}[h!]
    \centering
    \includegraphics[scale = 0.6]{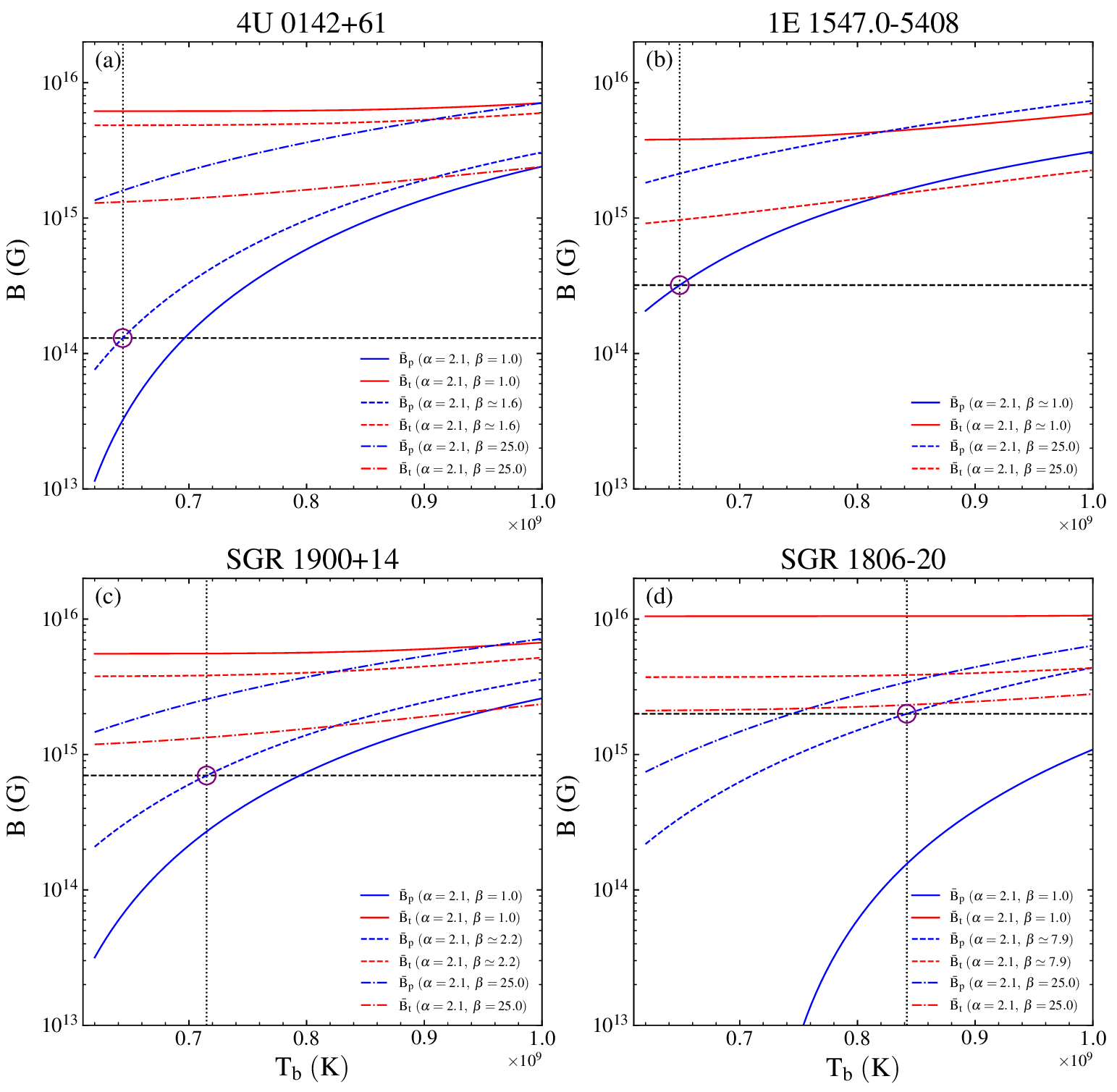}
    \caption{The volume-averaged strengths of the poloidal field $\bar{B}_{\rm p}$ (blue curves) and toroidal field $\bar{B}_{\rm t}$ (red curves) versus internal temperature $T_{\rm b}$ of the magnetar. Panels (a)--(d) respectively show the results of 4U 0142+61, 1E 1547.0-5408, SGR 1900+14, and SGR 1806-20. In the calculations, different values of $\beta$ are used (see the legends). The black dashed and dotted lines in each panel respectively represent the surface dipole field $B_{\rm s}$ and $T_{\rm b}$ inferred from surface thermal emission of the magnetar, as listed in Tab. \ref{tab1}. The purple open circle in each panel corresponds to the point that has $\bar{B}_{\rm p}=B_{\rm s}$, and $T_{\rm b}$ given in Tab. \ref{tab1}.} 
    \label{fig1}
\end{figure*}

\begin{figure*}[h!]
    \centering
    \includegraphics[scale=0.43]{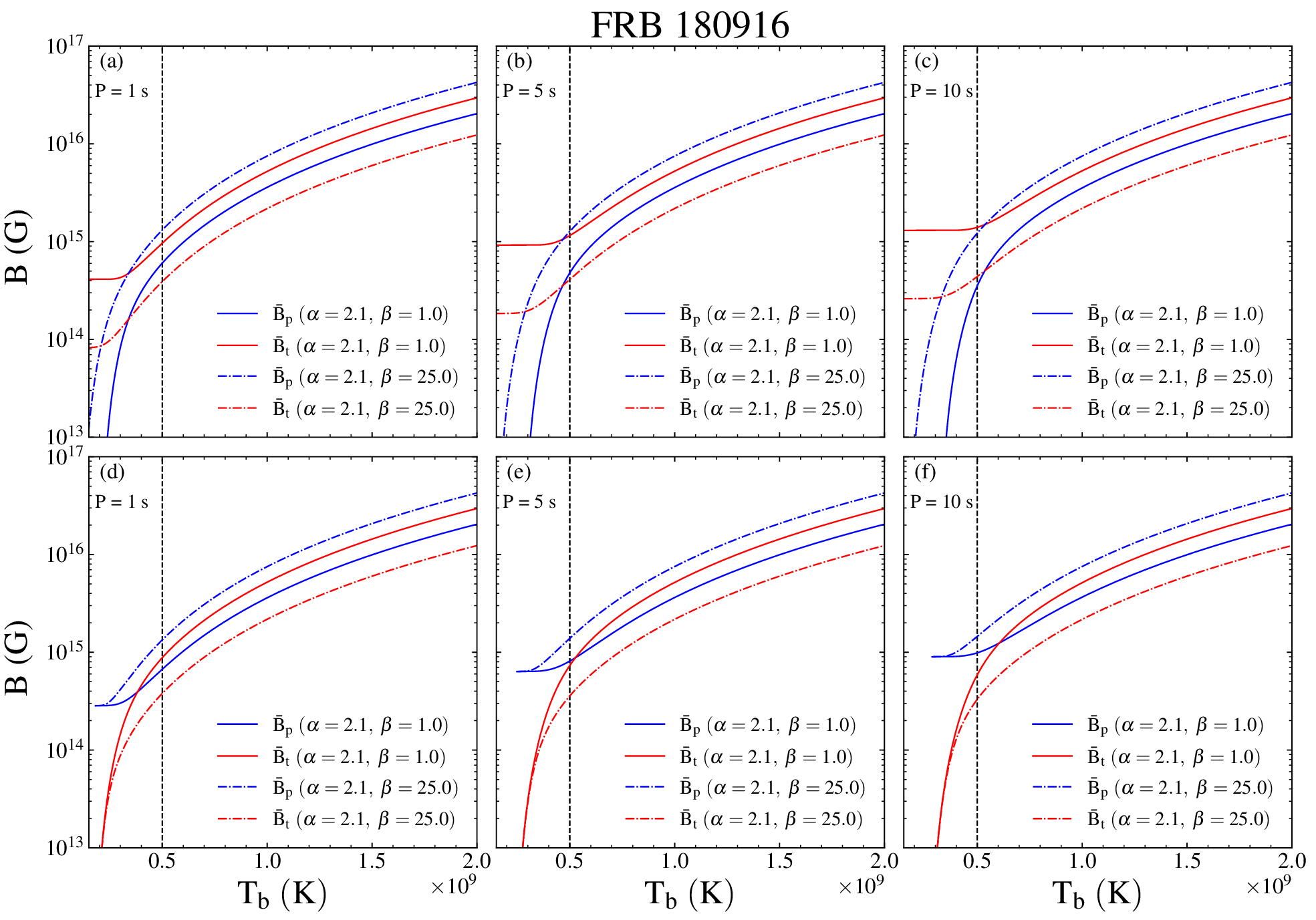}
    \caption{$\bar{B}_{\rm p}$ (blue curves) and $\bar{B}_{\rm t}$ (red curves) versus $T_{\rm b}$ of the putative host magnetar of FRB 180916 for various $\beta$ adopted (see the legends). The value of the spin period $P$ adopted for the host magnetar is labelled in each panel. In panels (a)--(c), the host magnetar is always assumed to have ellipticity $\epsilon_{\rm B}<0$, while in panels (d)--(f), $\epsilon_{\rm B}>0$ is always assumed. The black dashed line in each panel denotes a possible critical temperature for neutron superfluidity in the core, $T_{\rm c,core}=5\times10^8$ K \cite{2011PhRvL.106h1101P}.} 
    \label{fig2}
\end{figure*}


\begin{figure*}[h!]
    \centering
    \includegraphics[scale=0.43]{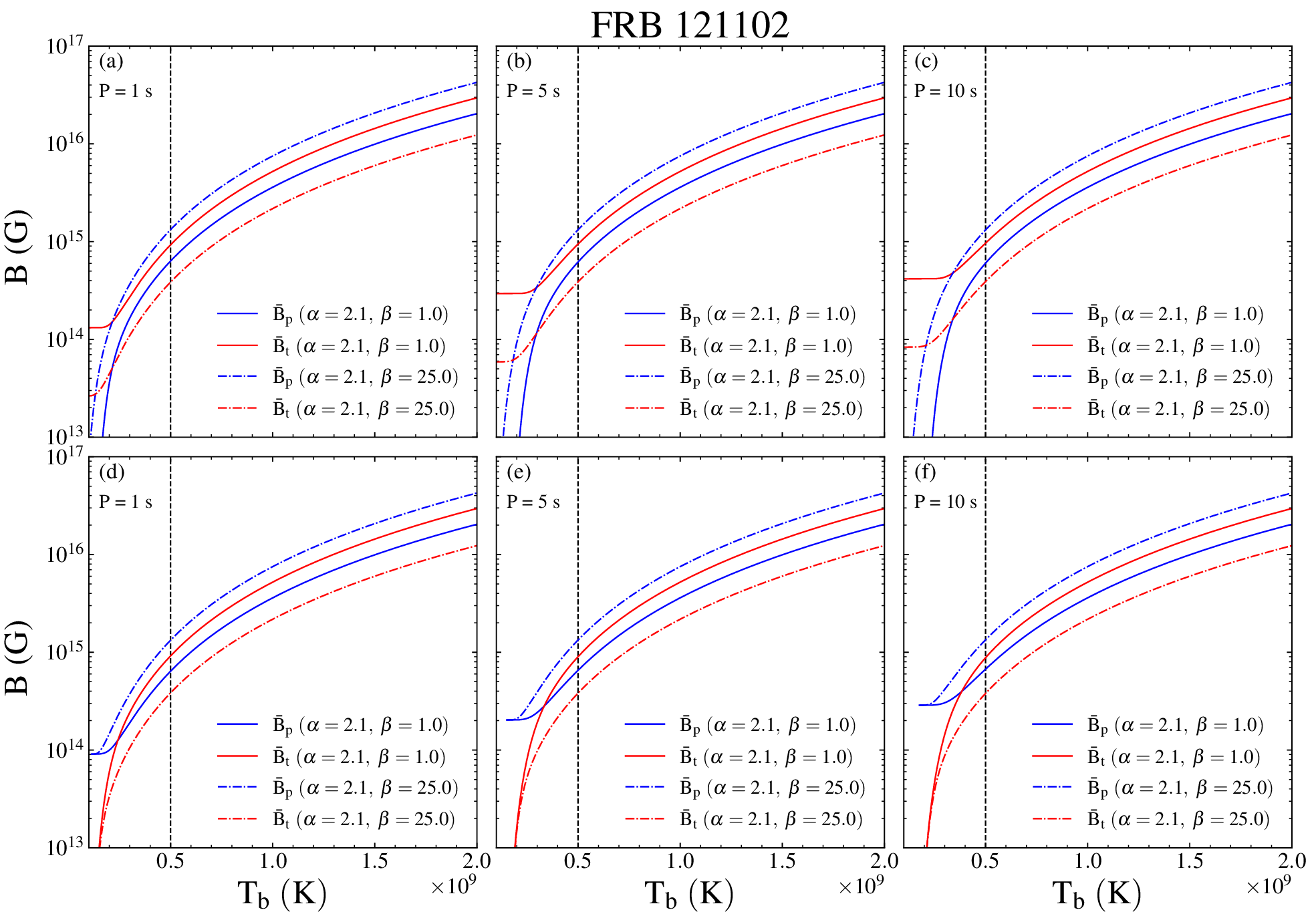}
    \caption{The same as Fig. \ref{fig2}, the results for the putative host magnetar of FRB 121102 are shown.}
    \label{fig3}
\end{figure*}

\section{Constraints on the poloidal and toroidal fields of the magnetars}\label{Sec IV}

By combining Eqs. (\ref{epsilonB}) and (\ref{Tb}), and substituting $\epsilon_{\rm B}$ listed in Tab. \ref{tab1} into Eq. (\ref{epsilonB}), and adopting a value for $\beta$, we can solve for $\bar{B}_{\rm p}$ and $\bar{B}_{\rm t}$ at specific $T_{\rm b}$. The curves of $\bar{B}_{\rm p}$ and $\bar{B}_{\rm t}$ versus $T_{\rm b}$ at various $\beta$ of the four magnetars as labeled are shown in Fig. \ref{fig1} (see the legends). The purple open circle in each panel corresponds to the magnetar that has $T_{\rm b}$ given in Tab. \ref{tab1}, and $\bar{B}_{\rm p}=B_{\rm s}$. As a general case, if the internal poloidal field smoothly connects with the surface dipole field of the magnetar, one possibly has $\bar{B}_{\rm p}=B_{\rm s}$. Therefore, to simultaneously satisfy the requirements of $\bar{B}_{\rm p}=B_{\rm s}$ and a reasonable $T_{\rm b}$ (see Tab. \ref{tab1}) inferred from the magnetar's surface thermal emission, the $\bar{B}_{\rm p}$--$T_{\rm b}$ curve should cross the purple open circle. For the four magnetars, this can be realized when appropriate values of $\beta$ are adopted. As shown in Fig. \ref{fig1}, we respectively have $\beta\simeq1.6$, 1.0, 2.2, 7.9 for 4U 0142+61, 1E 1547.0-5408, SGR 1900+14, and SGR 1806-20. Obviously, to meet the two requirements above, $\beta\gtrsim1$ are needed for these sources. In Tab. \ref{tab1}, we also present the ratios $\bar{B}_{\rm t}/\bar{B}_{\rm p}$ of these magnetars when the two requirements are satisfied, which are generally within the range $\sim2$--37. This is consistent with the range of the strength ratio of toroidal to poloidal fields of magnetars deduced from various observations and theoretical work \cite{2024EPJC...84.1043Y}. From Fig. \ref{fig1}, it is found that at this point these magnetars have $\bar{B}_{\rm p}\sim10^{14}$--$10^{15}$ G and $\bar{B}_{\rm t}\sim10^{15}$ G. 

Although $\beta<1$ is excluded for these sources as required by $\bar{B}_{\rm p}=B_{\rm s}$ and reasonable $T_{\rm b}$, a much larger $\beta$ (e.g., $\beta=25$) may still be reasonable. Fig. \ref{fig1} shows that when $\beta=25$ is used, one then has $\bar{B}_{\rm p}>B_{\rm s}$ for 4U 0142+61, 1E 1547.0-5408, and SGR 1900+14. Similarly, for SGR 1806-20 $\bar{B}_{\rm p}$ is also larger than $B_{\rm s}$ when $T_{\rm b}\gtrsim7.5\times10^8$ K. Theoretically, $\bar{B}_{\rm p}$ of a magnetar could indeed be larger than $B_{\rm s}$ because early fall-back accretion soon after the birth of the magnetar may bury its initially strong dipole field into the NS crust, resulting in a weaker dipole field whose strength is lower than its internal poloidal field \cite{1999A&A...345..847G,2024SCPMA..67.129513}. It should be pointed out that in this case the vertical black dotted line in each panel of Fig. \ref{fig1} should be shifted to the left to some extent. This is because for the same $T_{\rm s}$, the magnetar possibly has a lower $T_{\rm b}$ if it has a fully accreted envelope compared to the iron envelope considered in this work (see Fig. 7 of Ref. \cite{2003ApJ...594..404P}). Another interesting result is that when $\beta=25$ is adopted, $\bar{B}_{\rm p}$ of the four magnetars are generally larger than $\bar{B}_{\rm t}$. Nevertheless, these magnetars still have a prolate shape ($\epsilon_{\rm B}<0$). Therefore, in addition to the strengths of the poloidal and toroidal fields, the toroidal-field distribution (which is characterized by $\beta$) in the magnetar interior could also significantly affect its deformation.

Since the four magnetars are probably freely precessing, we therefore may set constraints on the critical temperature for $^3P_2$ neutron superfluidity by using $T_{\rm b}>T_{\rm c,core}$. For the four magnetars, the critical temperature roughly needs to satisfy $T_{\rm c,core}<6.4$--$8.4\times10^8$ K. The most stringent constraint comes from 4U 0142+61, which requires $T_{\rm c,core}<6.4\times10^8$ K. This constraint is consistent with the results from modeling the cooling curve of the NS in Cassiopeia A \cite{2011PhRvL.106h1101P} and other theoretical calculations \cite{2016PhRvC..94b5802D,2021MNRAS.500.1505D}.  

The putative host magnetars of FRBs 180916 and 121102 respectively have $P_{\rm p}\simeq16.35$ and 160 d \cite{2020Natur.582..351C,2021MNRAS.500..448C}, thus their $|\epsilon_{\rm B}|$ can be derived without knowledge of the sign of $\epsilon_{\rm B}$ by assuming $P=1$, $5$, and $10$ s. Using the same method as introduced at the beginning of this section, we can obtain the curves of $\bar{B}_{\rm p}$ and $\bar{B}_{\rm t}$ versus $T_{\rm b}$ at various $\beta$ (see the legends) for the magnetars of FRBs 180916 (Fig. \ref{fig2}) and 121102 (Fig. \ref{fig3}). The upper panels of Figs. \ref{fig2} and \ref{fig3} show the results when these magnetars have $\epsilon_{\rm B}<0$, while the lower panels present the results of $\epsilon_{\rm B}>0$. The left, middle, and right panels of Figs. \ref{fig2} and \ref{fig3} respectively exhibit the results when the two magnetars have $P=1$, $5$, and $10$ s as labeled. The black dashed line in each panel shows a possible critical temperature $T_{\rm c,core}=5\times10^8$ K for neutron superfluidity as deduced from the cooling of the NS in Cassiopeia A \cite{2011PhRvL.106h1101P}. To account for the periodicities of FRBs 180916 and 121102 in the freely-precessing-magnetar scenario, the host magnetars should have internal temperatures $T_{\rm b}>T_{\rm c,core}$. Hence, only the parameter space on the right of the black dashed line in each panel is available. For instance, when $\epsilon_{\rm B}<0$, $\beta=1$, and $P=1$ s are adopted, the host magnetar of FRB 180916 should have $\bar{B}_{\rm p}\gtrsim6\times10^{14}$ G and $\bar{B}_{\rm t}\gtrsim10^{15}$ G. However, by using $\epsilon_{\rm B}>0$, $\beta=1$, and $P=1$ s, we have $\bar{B}_{\rm p}\gtrsim7\times10^{14}$ G and $\bar{B}_{\rm t}\gtrsim9\times10^{14}$ G. It is found that different signs of $\epsilon_{\rm B}$ can affect the constraints on $\bar{B}_{\rm p}$ and $\bar{B}_{\rm t}$ of the host magnetar of FRB 180916 by a facor of $\sim1$--$3$, while different values of $P$ can only change the limits by a facotr of $\sim1$--$2$. When the host magnetar of FRB 180916 has a spin period $P\sim1$--$10$ s and $\beta=1$ is used, the following constraints on its internal magnetic fields can be set 
\begin{eqnarray}
\left\{ \begin{aligned}
         &\bar{B}_{\rm p}\gtrsim4\mbox{--}6\times10^{14}~{\rm G},~~~~~~~~~~~~~~{\rm for}~\epsilon_{\rm B}<0, \\
                &\bar{B}_{\rm t}\gtrsim10^{15}~{\rm G},~~~~~~~~~~~~~~~~~~~~~~~~~{\rm for}~\epsilon_{\rm B}<0, \\
         &\bar{B}_{\rm p}\gtrsim7\mbox{--}10\times10^{14}~{\rm G},~~~~~~~~~~~~{\rm for}~\epsilon_{\rm B}>0, \\
         &\bar{B}_{\rm t}\gtrsim6\mbox{--}9\times10^{14}~{\rm G},~~~~~~~~~~~~~~~{\rm for}~\epsilon_{\rm B}>0.
                          \end{aligned} \right.
                          \label{Eq7}
\end{eqnarray}
Using $\beta=25$ and $\epsilon_{\rm B}<0$, the constraints on the internal magnetic fields are $\bar{B}_{\rm p}\gtrsim10^{15}$ G and $\bar{B}_{\rm t}\gtrsim4\times10^{14}$ G if the host magnetar has $P=1$ s (see Fig. \ref{fig2}). This suggests that if $\beta$ is large enough, $\bar{B}_{\rm p}$ will be larger than $\bar{B}_{\rm t}$ though the host magnetar still has $\epsilon_{\rm B}<0$, as found in 4U 0142+61. Similarly, assuming $P\sim1$--$10$ s and using $\beta=25$, the constraints on $\bar{B}_{\rm p}$ and $\bar{B}_{\rm t}$ of the host magnetar of FRB 180916 are
\begin{eqnarray}
\left\{ \begin{aligned}
         &\bar{B}_{\rm p}\gtrsim10^{15}~{\rm G},~~~~~~~~~~~~~~~~~~~~~~~{\rm for}~\epsilon_{\rm B}<0~{\rm and}~\epsilon_{\rm B}>0, \\
                &\bar{B}_{\rm t}\gtrsim4\times10^{14}~{\rm G},~~~~~~~~~~~~~~~~~{\rm for}~\epsilon_{\rm B}<0~{\rm and}~\epsilon_{\rm B}>0.
                          \end{aligned} \right.
                          \label{Eq8}
\end{eqnarray}

From Fig. \ref{fig3}, the constraints on the internal fields of the host magnetar of FRB 121102 can also be derived. When $P\sim1$--$10$ s and $\beta=1$ are used, we have
\begin{eqnarray}
\left\{ \begin{aligned}
         &\bar{B}_{\rm p}\gtrsim6\times10^{14}~{\rm G},~~~~~~~~~~~~~~~~{\rm for}~\epsilon_{\rm B}<0~{\rm and}~\epsilon_{\rm B}>0, \\
                &\bar{B}_{\rm t}\gtrsim9\times10^{14}~{\rm G},~~~~~~~~~~~~~~~~~{\rm for}~\epsilon_{\rm B}<0~{\rm and}~\epsilon_{\rm B}>0.
                          \end{aligned} \right.
                          \label{Eq9}
\end{eqnarray}
While keeping $P$ unchanged, if $\beta=25$ is used, the internal fields of this host magnetar should satisfy 
\begin{eqnarray}
\left\{ \begin{aligned}
         &\bar{B}_{\rm p}\gtrsim10^{15}~{\rm G},~~~~~~~~~~~~~~~~~~~~~~~{\rm for}~\epsilon_{\rm B}<0~{\rm and}~\epsilon_{\rm B}>0, \\
                &\bar{B}_{\rm t}\gtrsim4\times10^{14}~{\rm G},~~~~~~~~~~~~~~~~~{\rm for}~\epsilon_{\rm B}<0~{\rm and}~\epsilon_{\rm B}>0.
                          \end{aligned} \right.
                          \label{Eq10}
\end{eqnarray}

\section{Conclusions and discussions}\label{Sec V}

In this work, the modulation periods observed in the hard X-ray emissions of the magnetars 4U 0142+61, 1E 1547.0-5408, SGR 1900+14, and SGR 1806-20, as well as the periodicities of FRBs 180916 and 121102 are all attributed to the free-precession periods of the (host) magnetars. Since the spin periods and surface temperatures of the first four magnetars are available, combining these quantities with the free-precession periods, we can estimate their internal magnetic fields, which are generally hard to be probed with other methods. We may also set constraints on the internal fields of the putative host magnetars of FRBs 180916 and 121102 by assuming that their spin periods are $P\sim1$--$10$ s and using the critical temperature of neutron superfluidity in the core $T_{\rm c,core}=5\times10^8$ K \cite{2011PhRvL.106h1101P}. In our calculations, an expression which explicitly shows the contributions of internal poloidal and toroidal fields is used for the magnetically-induced ellipticity $\epsilon_{\rm B}$. Moreover, we also take into account the effect of toroidal-field distribution on $\epsilon_{\rm B}$, which is characterized by the coefficient $\beta$. 

To account for the surface thermal emissions of 4U 0142+61, 1E 1547.0-5408, SGR 1900+14, and SGR 1806-20, and simultaneously require that their internal poloidal fields smoothly connect with the external dipole fields ($\bar{B}_{\rm p}=B_{\rm s}$), $\beta\gtrsim1$ are needed for these sources and their toroidal fields $\bar{B}_{\rm t}$ are $\sim2$--$37$ times larger than $\bar{B}_{\rm p}$. Therefore, to satisfy the two requirements, the poloidal and toroidal fields of the four magnetars are inferred to be $\bar{B}_{\rm p}\sim10^{14}$--$10^{15}$ G and $\bar{B}_{\rm t}\sim10^{15}$ G. We also find that a much larger $\beta$ (e.g., $\beta=25$) could also be reasonable since the initially strong dipole fields were possibly submerged into the NS crust due to fall-back accretion \cite{1999A&A...345..847G,2024SCPMA..67.129513}. Interestingly, these magnetars would generally have $\bar{B}_{\rm p}>\bar{B}_{\rm t}$ in this case though their ellipticities are $\epsilon_{\rm B}<0$. We may also constrain $T_{\rm c,core}$ considering that the four magnetars are probably precessing. From 4U 0142+61, we could obtain $T_{\rm c,core}<6.4\times10^8$ K, which is the most stringent constraint for the sample investigated in this work. 

The internal fields of the host magnetars of FRBs 180916 and 121102 are relatively difficult to be determined due to the lack of spin periods and surface thermal emissions observed. Our results show that the ranges of $\bar{B}_{\rm p}$ and $\bar{B}_{\rm t}$ of the two host magnetars could be affected by $P$, $\epsilon_{\rm B}$, and $\beta$ adopted. The constraints on their $\bar{B}_{\rm p}$ and $\bar{B}_{\rm t}$ of  are given in Eqs. (\ref{Eq7})--(\ref{Eq10}), which can simply be summarized as $\bar{B}_{\rm p}\gtrsim10^{14}$--$10^{15}$ G and $\bar{B}_{\rm t}\gtrsim10^{14}$--$10^{15}$ G. The results thus indicate that the internal fields of the two host magnetars possibly have similar strengths as that of 4U 0142+61, 1E 1547.0-5408, SGR 1900+14, and SGR 1806-20, though their precession periods are remarkably different.

In previous work \cite{2014PhRvL.112q1102M,2021MNRAS.502.2266M,2021ApJ...923...63M,2024PASJ...76..688M,2020ApJ...895L..30L,2022ApJ...928...53W}, the internal fields of the four magnetars, and host magnetars of the two FRBs studied in this work were inferred to be $\sim10^{15}$--$10^{16}$ G. Compared to their results, our constraints on the internal fields of these sources are about a few to ten times smaller. This could be attributed to different forms of $\epsilon_{\rm B}$ adopted here and in previous work since a relatively large $\beta$ (see Eq. (\ref{epsilonB})) can reduce the required strength of toroidal field in deforming the NS. This, to some extent, may alleviate the contradiction between large toroidal fields needed to produce large enough $\epsilon_{\rm B}$ and the requisite of magnetoelastic equilibrium in the crust \cite{2024MNRAS.527.2297K}. Nevertheless, the existence of a stable magnetic field configuration that could give rise to $\epsilon_{\rm B}$ described by Eq. (\ref{epsilonB}) remains to be tested by magnetohydrodynamics simulations \cite{2009MNRAS.395.2162L,2009MNRAS.397..913C,2013PhRvD..88j3005L}. Moreover, the complicated periodicities of FRB 121102 \cite{2024ApJ...969...23L} possibly indicates that substantial modifications to the precessing-magnetar scenario are needed when accounting for its periodicities.    
 
\begin{acknowledgments}
We gratefully thank the anonymous referee for helpful comments. This work is supported by the National Natural Science Foundation of China (Grant No. 12033001, No. 12473039, and No. 12003009), and the National SKA program of China (Grant No. 2020SKA0120300).
\end{acknowledgments}

\bibliographystyle{apsrev4-2}
\bibliography{reference}

\end{document}